\documentclass[11pt,twoside]{article}
\usepackage{newpasp}
\usepackage{epsf}
\newcommand\etal{{\it et al.}}

\index{Haynes, M.P.}
\index{replace this with author2's surname, initials}
\index{replace this with author3's surname, initials}

\begin{document}

\title{Kinematic Evidence of Minor Mergers in Normal Sa Galaxies}
\author{Martha P. Haynes}
\affil{Center for Radiophysics and Space Research, Space Sciences Building,
Cornell University, Ithaca, NY 14853, USA}


\begin{abstract}
A detailed study of nearby, morphologically normal
Sa galaxies reveals that about half show some spectrocopic
evidence for kinematically distinct components despite their
undisturbed optical appearance. In eight of nine objects mapped 
in the 21~cm HI line, the HI distribution extends far outside the 
optical disk and warps of the HI disk are complex. The multiwavelength 
evidence can be interpreted in terms of the kinematic ``memory'' of 
past minor mergers in objects that otherwise exhibit no morphological 
signs of interaction.
 
\end{abstract}




\section{Introduction}

Sa galaxies are the earliest Hubble types exhibiting clear evidence of 
spiral structure. But while the small pitch angle of the arms
in Sa's is distinctive, tightly wound spirals show a wide
range of bulge size and current star formation rate, the other two criteria
used to distinguish among the spiral types. Thus, the Sa class is
heterogeneous, including gas-rich and gas-poor disks and large and 
small--bulged systems. Furthermore, in contrast to their later spiral counterparts,
Sa's typically occupy higher density environments and
require little or no dark matter within their optical disks.

The origin of the heterogeneity of the Sa class has been the focus of
a detailed study of the morphology, environment, kinematics
and dynamics of a sample of nearby, undisturbed and relatively
isolated Sa galaxies. Broad--band optical imaging observations provide luminosity
profiles, colors and measures of morphological asymmetry in the stellar 
distribution. Long--slit spectroscopy along both the major and minor axes
provide details of the kinematics of both stellar and ionized gas components.
Narrow--band H$\alpha$ imaging identifies the sites of current massive
star formation, while HI synthesis imaging provides measures of both
morphological and dynamical asymmetry in the gas disk and details of the
dynamics which can be used to trace the mass. The body of this work 
is presented in Jore (1997), Kornreich (2001), Jore \etal ~(1996), Haynes \etal
~(2000), Jore \etal ~(2000) and Kornreich \etal ~(2000).



\section{Counterrotation in NGC~4138}

Jore \etal ~(1996) presented results for the surprising case of NGC~4138,
the first Sa we studied, specifically chosen for its relative isolation and
morphological ``boring''--ness. NGC~4138 is a smooth, undisturbed, small-bulged 
Sa, classified by the small pitch angle of its arms. An outlying member of the 
Ursa Major cluster, its nearest neighbor lies more than 9 times further 
than the observed HI extent. As evident in Figure 1, the HI gas is extended,
with R$_{HI}$/R$_{opt}$ = 2.2, and the HI surface density is very low throughout.
The HI rotation curve can be traced to 16 disk scale lengths, and while
the {\it observed} HI rotation curve declines by 80 km~s$^{-1}$ beyond the
optical edge, a significant warp is evident from the bending of the 
minor axis velocity contours, making the derivation of the mass
distribution ambiguous. Nonetheless, at least 50\% of the total mass within
the optical radius is dark. Most curiously, however, all of
the gas, as traced both by the HI 
and the optical emission lines, counterrotates with respect to the
primary stellar component; a secondary stellar population, kinematically
coupled to the gas, is also visible. The amplitude of the cross-correlation function 
of the stellar lines coincides spatially with an H$\alpha$ ring, previously noticed by
Pogge \& Eskridge ~(1987) and also shown in Figure 1, at a radius of 
$\sim$ 20\arcsec. The two stellar components are characterized by distinct 
velocity dispersions, implying separate origins. Thakar \etal ~(1997)
showed that the extended counter--rotating component might be the result
either of the accretion of a gas--rich dwarf or retrograde primordial gas
infall.

\begin{figure}
\plotfiddle{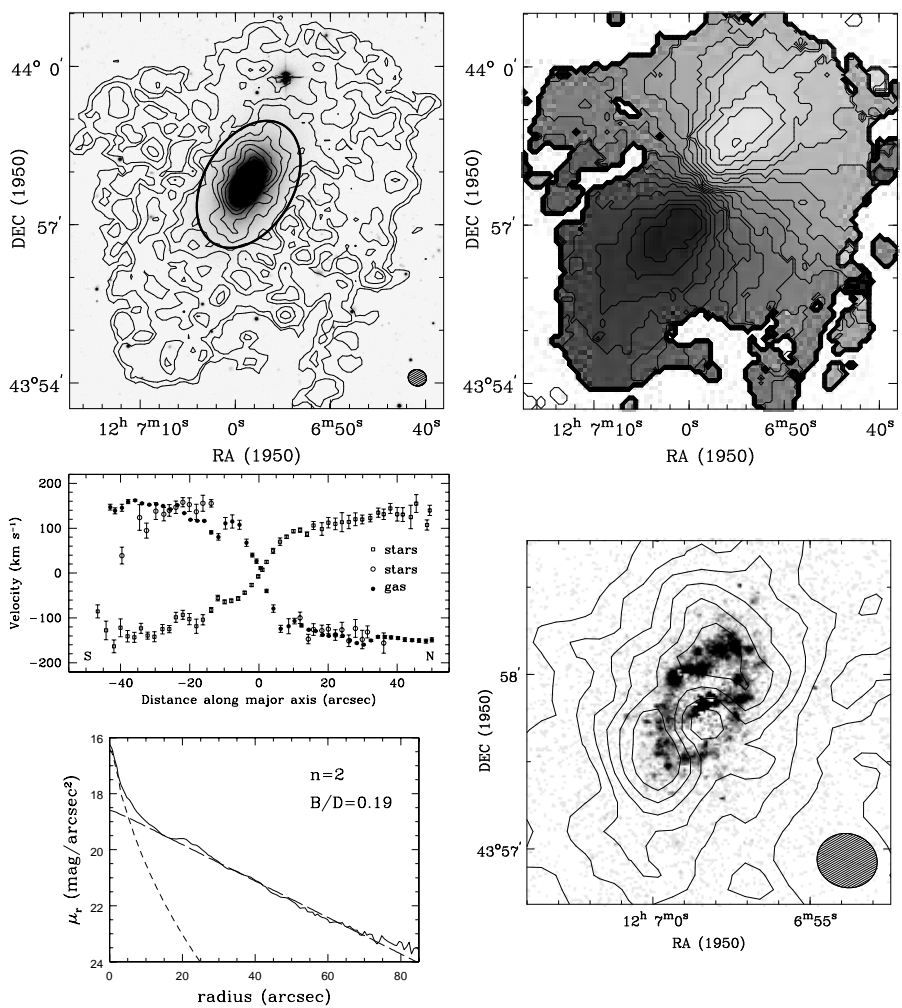}{4.0in}{0}{90}{90}{-252}{-250}
\vskip 25pt
\caption[]{A montage of results for NGC~4138. Upper left: HI column 
density distribution superposed on $V$ band image. The superposed ellipse
has dimensions D25$\times$d25 oriented along the optical major axis. Center left:
Observed velocities derived from the optical long--slit spectra
for stellar (MgIb) absorption and ionized gas emission ([NII]) lines 
along the major axis. 
Lower left: $R$ band surface brightness profile and the best--fit
decomposition into disk and a generalized exponential $n$=2 bulge.
Upper right: HI velocity field. Lower right: Inner HI column density
distribution superposed on the H$\alpha$
image. After Jore \etal ~(1996) and Haynes \etal ~(2000).}
\end{figure}

\section{More Evidence of Minor Mergers}

Of the overall sample of 20 Sa galaxies studied by Jore (1997), nearly half show
some spectroscopic evidence of kinematic peculiarity. The degree
and circumstances of the distinct kinematics vary from complete counterrotation of 
all of the gas from all/most of the stars (NGC~3626, NGC~4138) to nuclear gas disks 
decoupled from the stars (NGC~5854) to anomalous velocity central gas components
(NGC~3623, NGC~3900, NGC~4772). The HI distribution in nine Sa's was mapped with 
the VLA, and of those, all but NGC~3623, known to be a member of the interacting 
Leo Triplet, show R$_{HI}$/R$_{opt} \sim$ 2. Rings are prominent 
features in both the optical light and in the distributions of ionized and 
neutral gas. While the HI velocity fields are dominated by circular rotation, 
the HI disks are significantly warped.
In general, the HI surface density is very low and the outer HI is patchy
and asymmetric (NGC~3900, NGC~4138) or found in a distinct ring, exterior to
the optical edge (NGC~3626, NGC~4772, NGC~5448). While the overall HI
velocity fields are dominated by circular motions, warps are
suggested in the outer regions by bending of the minor axis isovelocity
contours (NGC~1169, NGC~4138) and/or systematic shifts in position angle of
the inner and outer rings (NGC~3626, NGC~4772). In the eight cases where the
HI is extended, significant dark matter halos are required; in the other case 
(NGC~3623), the mass of the halo cannot be constrained by the detected rotation
curve.

Kornreich \etal~ (2000) apply quantitative measures of morphological and dynamical
asymmetry to the HI synthesis maps. While the stellar light distribution is
axisymmetric, significant deviations from symmetry were found in both the HI
morphology and dynamics. The complex kinematics seen in both the the optical long--slit
spectroscopic data and the HI velocity fields is also evident in the warping of the HI
disks. With the exception of the kinematically-normal SBa NGC~1169, none
of these galaxies shows the simple warping behavior seen in the Briggs' (1990) sample.
Rather, the warping observed in most of the Sa galaxies is more complex and of order
15$^\circ$, as predicted by the simulations of minor mergers presented by 
Quinn \etal~ (1993) and Hernquist \& Mihos (1995).

While Sa's tend to be found in higher density environments, it may be very important
that the present sample consists principally of relatively isolated, unbarred and 
undisturbed Sa galaxies. Although all show tightly wound arms, they span a range of 
bulge--to--disk ratios (B/D). Giuricin \etal~ (1995) have concluded
that the observed tendency of galaxies to have greater B/D in higher
density regions is simply reflective of morphological segregation. To
try to see if B/D might correlate with environment
{\it within the Sa class}, we have
applied the 2-D decomposition scheme of 
Moriondo \etal~ (1998) to derive $I$ band B/D for 
a nearly-complete sample of northern Sa's fitting a generalized
exponential to the bulge light,
$I_b(r) = I_{e}exp[-(r/r_e)^{1/n}],$ where $n$ is the shape index. 
In on-going work, we find marginal evidence that Sa's with 
close neighbors show a larger range of B/D than ones with no
near neighbors (Hagemann 2000) and that $n$ may vary with environment.

While the current sample is small and consists of relatively isolated
objects, their optical morphology bears no hint of the disturbances responsible for the
complex kinematics evident in half the galaxies. Thus, minor mergers may play
an important role in producing the heterogeneity of the Sa class through the
triggering of starbursts, bulge--building or the dislocation of disk gas, but
the evidence of such minor events is quickly lost from the morphological appearance
of the stellar distribution as evident in optical images. Morphological criteria
alone thus may seriously underestimate the rate of minor-to-moderate mergers.

\acknowledgements
I would know little about minor mergers in Sa galaxies were it not
for my Cornell students Liz Barrett (M.Engr. 1999), Katrin Hagemann (M.S. 2000),
Katie Jore (Ph.D. 1997), and Dave Kornreich (Ph.D. 2001) whose hard work
and adviser--tolerance I gratefully acknowledge.
This work has been supported by NSF grants AST--9023450, 
AST--9528860, and AST-9900695.

\end{document}